 \documentclass[11pt,twocolumn]{article}

 \pdfoutput=1

 \usepackage[english]{babel}
\usepackage{graphicx} 
\usepackage{amsmath}

\usepackage{enumitem}

\usepackage{float}
\usepackage{ifthen}
\usepackage{hyperref}

\usepackage{titlesec}

\titlespacing*{\section}{0pt}{0.2\baselineskip}{\baselineskip}

\begin{document}

\pagestyle{empty}


\twocolumn[
\begin{center}
{\bf \huge 
The mixed-phase version of moist-air entropy.
}\\
\vspace*{3mm}
{\Large \bf by Pascal Marquet } {\Large (WGNE Blue-Book 2016)}. \\
\vspace*{2mm}
{M\'et\'eo-France. CNRM/GMAP.
 Toulouse. France.}
{\it E-mail: pascal.marquet@meteo.fr} \\
\vspace*{1mm}
\end{center}
]





 \section{\underline{\Large Motivations}} 
\vspace{-4mm}

The specific (per unit mass of {\it moist-air\/} entropy is defined in Marquet (2011, M11) by $s  =  s_{ref}  + c_{pd} \: \ln(\theta_{s})$, where $s_{ref}$ and $c_{pd}$ are two constants.
The first- and second-order approximations $({\theta}_{s})_1$ and $({\theta}_{s})_2$ of the moist-air entropy potential temperature ${\theta}_{s}$ have been more recently derived in Marquet (2015, M15).

The aim of this note is to derive the {\it mixed-phase\/} version of $\theta_s$, $({\theta}_{s})_1$ and $({\theta}_{s})_2$, namely if liquid water and ice are allowed to coexist, with possible under- or super-saturations, with possible supercooled water and with possible different temperatures for dry air and water vapour, on the one hand, condensed water and ice, on the other hand.

 \section{\underline{\Large The mixed-phase definition of $\theta_s$}} 
\label{section2}
\vspace{-4mm}

The specific  (per unit mass of {\it moist-air\/}) entropy given by (B.1) in M11 is equal to the sum 
\vspace*{-2mm}
\begin{align}
s & \: = \; q_d \: s_d \: + \: q_v \: s_v \: + \: q_l \: s_l \: + \: q_i \: s_i 
\label{eq_1} \: ,
\end{align}

\vspace*{-3mm}
\noindent where specific contents in dry-air, water vapor, liquid water and ice ($q_d$, $q_v$, $q_l$, $q_i$) act as weighting factors.
The common temperature $T$ for the dry air and water vapour entropies ($s_d$, $s_v$) is possibly different from those $T_l$ or $T_i$ for liquid water or ice entropies ($s_l$, $s_i$), respectively.

Without lost of generality, the moist-air entropy given by (\ref{eq_1}) can be rewritten in a way similar to (B.2) in M11, leading to
\vspace*{-2mm}
\begin{align}
s & \: = \; q_d \: s_d \: + \: q_t \: s_v 
       \: + \: q_l \: (s_l^{\ast}-s_v) 
       \: + \: q_i \: (s_i^{\ast}-s_v) 
  \nonumber \\
  & \; \; 
       \: + \: q_l \: (s_l - s_l^{\ast}) 
       \: + \: q_i \: (s_i - s_i^{\ast})
\label{eq_2} \: ,
\end{align}

\vspace*{-3mm}
\noindent where $q_t = q_v + q_l +q_i$ is the total water content.

The first difference from the result derived in M11 is due to $s_l$ and $s_i$ which must be computed in the second line of (\ref{eq_2}) at temperatures $T_l$ and $T_i$, respectively, whereas $s_l^{\ast}$ and $s_i^{\ast}$ are computed at the common temperature $T$ for the two gaseous species.
The second line of (\ref{eq_2}) can thus be computed with
$s_l - s_l^{\ast} = c_l \: \ln(T_l/T)$
and
$s_i - s_i^{\ast} = c_i \: \ln(T_i/T)$,
where the reference entropies $(s_l)_r$ and $(s_i)_r$  have no impact.

The other difference concerns the bracketed terms in (B.7) in M11, namely the term
$R_v \: [ \: q_l \: \ln({H}_l) \: + \: q_l \: \ln({H}_i) \: ]$, where 
${H}_l = e/e_{sl}$ and ${H}_i = e/e_{si}$
are the relative humidities with respect to liquid water and ice, respectively. 
These bracketed terms no longer cancel out if liquid water and ice are allowed to coexist, and/or with possible under- or super-saturations.

These differences with respect to non-mixed phase results of M11 lead to the following mixed-phase generalisation of $\theta_s$~:
\vspace*{-1mm}
\begin{align}
  {\theta}_{s} 
   & = \: \left[ \: \theta \;
    \exp\! \left( - \frac{L_v \: q_l + L_s \: q_i}{c_{pd} \: T} \right)
          \: \right] \:
    \exp\! \left(  \Lambda_r \: q_t  \right)
\nonumber \\
   &
        \left( \frac{T}{T_r}\right)^{\!\!\lambda \,q_t}
 \!  \! \left( \frac{p}{p_r}\right)^{\!\!-\kappa \,\delta \,q_t}
 \!  \! \left( \frac{r_r}{r_v} \right)^{\!\!\gamma\,q_t}
      \frac{(1\!+\!\eta\,r_v)^{\,\kappa \, (1+\,\delta \,q_t)}}
           {(1\!+\!\eta\,r_r)^{\,\kappa \,\delta \,q_t}}
\nonumber \\
   & \; {\left( H_l \right)}^{\, \gamma \, q_l}
     \; {\left( H_i \right)}^{\, \gamma \, q_i}
     \; {\left( \frac{T_l}{T} \right)}^{\! c_l\, q_l/c_{pd}}
     \; {\left( \frac{T_i}{T} \right)}^{\! c_i\, q_i/c_{pd}}
\label{eq_3} \: .
\end{align}

\vspace*{-3mm}
\noindent The bracketed terms in the first line of (\ref{eq_3}) is the ice-liquid version of the Betts' potential temperature $\theta_l$, where the latent heats $L_v$ and $L_s$ depends on $T$.
The whole first line of (\ref{eq_3}), including the term $\exp ( \Lambda_r \: q_t )$ which depends on the Third-Law reference values $(s_v)_r$ and $(s_d)_r$, forms the first-order approximation $({\theta}_{s})_1$.
Some of the terms in the second line of (\ref{eq_3}) are used in M15 to derive the second-order approximations $({\theta}_{s})_2$.

The third line of (\ref{eq_3}) is made of the four new {\it mixed-phase correction terms\/}.
These terms are clearly equal to unity for the non-mixed phase conditions retained in M11, namely if 
$T_l=T_i=T$, 
$H_l=1$ for $q_l \neq 0$
and
$H_i=1$ for $q_i \neq 0$.

\begin{figure}[hbt]
\centering
\includegraphics[width=0.505\linewidth]{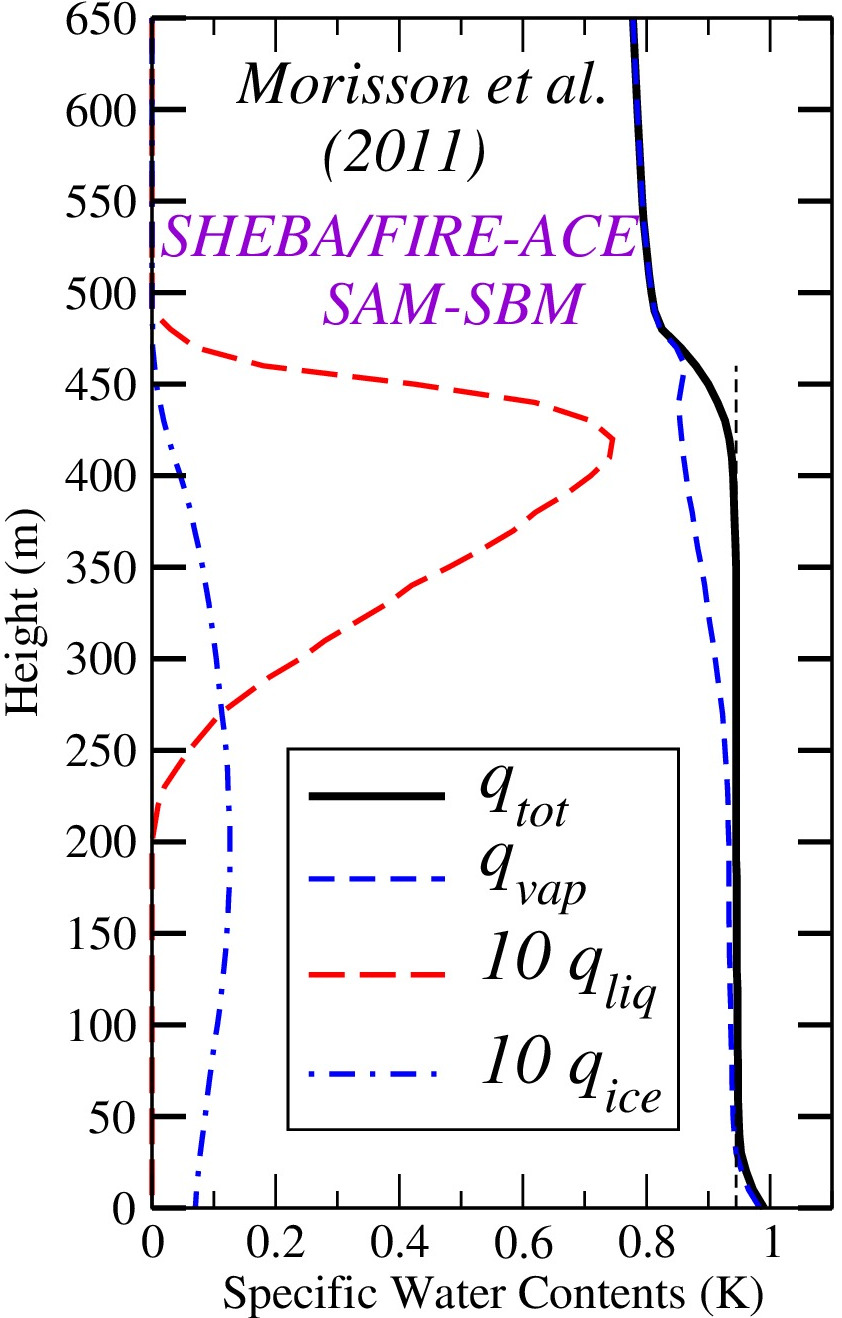}
\includegraphics[width=0.475\linewidth]{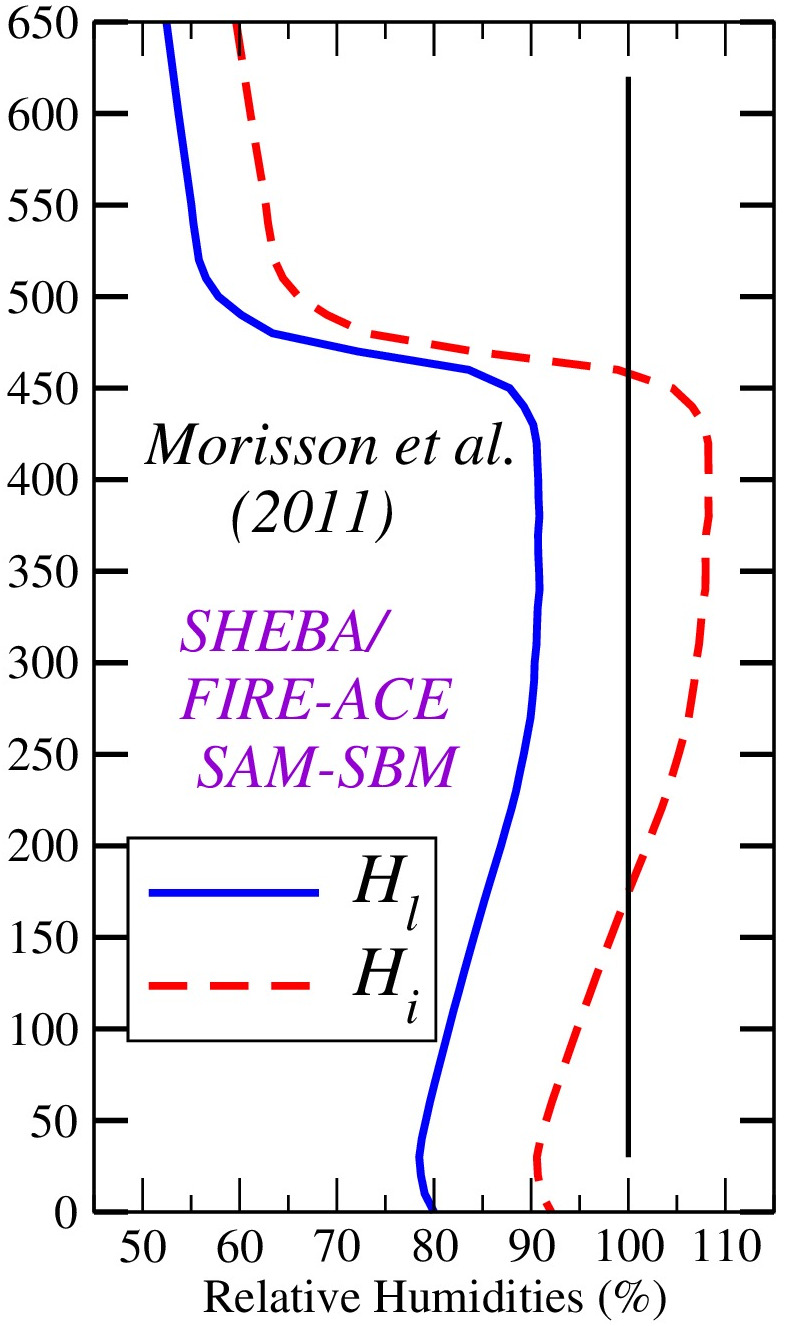}
\vspace{-3mm}
\caption{\small \it 
The vertical profile of water species contents and relative humidities corresponding to the yellow curve (SAM-SBM) in Figure~7 of Morisson {\it et al.\/} (2011).
\label{fig_1}}
\end{figure}

 \section{\underline{\Large Some Numerical results}} 
\label{section3}
\vspace{-4mm}

The impact of the two new mixed-phase terms ${( H_l )}^{\, \gamma \, q_l}$ and  ${( H_i )}^{\, \gamma \, q_i}$ in (\ref{eq_3}) are evaluated by using SHEBA/FIRE-ACE vertical profiles for ($\theta_l$, $q_t$, $q_l$, $q_i$) depicted in Figure~7 of Morisson {\it et al.\/} (2011).

The profiles of ($q_t$, $q_v$, $q_l$, $q_i$) and ($H_l$, $H_i$) are shown in Fig.\ref{fig_1}.
The contents in liquid water and ice are small (mind the factor $10$!), but they are associated with relative humidities mostly different from $100$~\%.
One may thus expect the factors ${( H_l )}^{\, \gamma \, q_l}$ and  ${( H_i )}^{\, \gamma \, q_i}$ to be (slightly) different from unity.
\begin{figure}[hbt]
\centering
\includegraphics[width=0.47\linewidth]{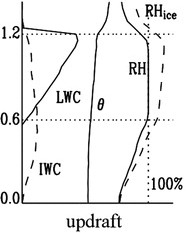}
\includegraphics[width=0.44\linewidth]{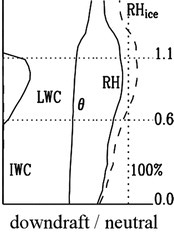}
\vspace{-3mm}
\caption{\small \it 
The conceptual model depicted in Shupe {\it et al.\/} (2008) showing typical values for water species contents, $\theta$ and relative humidities in autumn Arctic mixed-phase stratiform clouds (for updraft and downdraft regions).
\label{fig_2}}
\end{figure}
The vertical profiles $H_l(z)$ and $H_i(z)$ shown in Fig.\ref{fig_1} are similar to those described in Fig.\ref{fig_2} for Arctic mixed-phase clouds, with liquid and ice water content typical of updrafts and relative humidities typical of downdrafts.

\begin{figure}[hbt]
\centering
\includegraphics[width=0.518\linewidth]{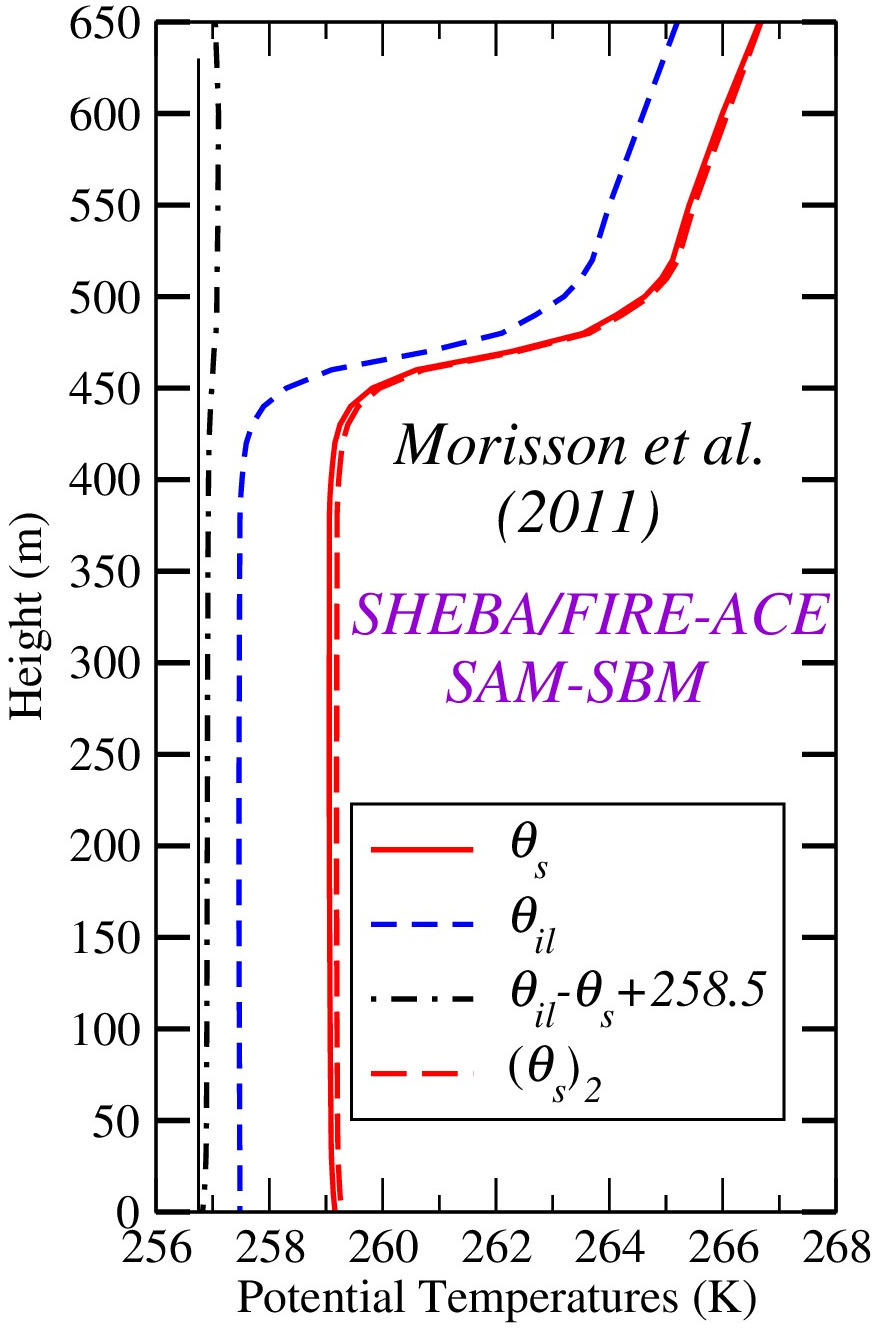}
\includegraphics[width=0.462\linewidth]{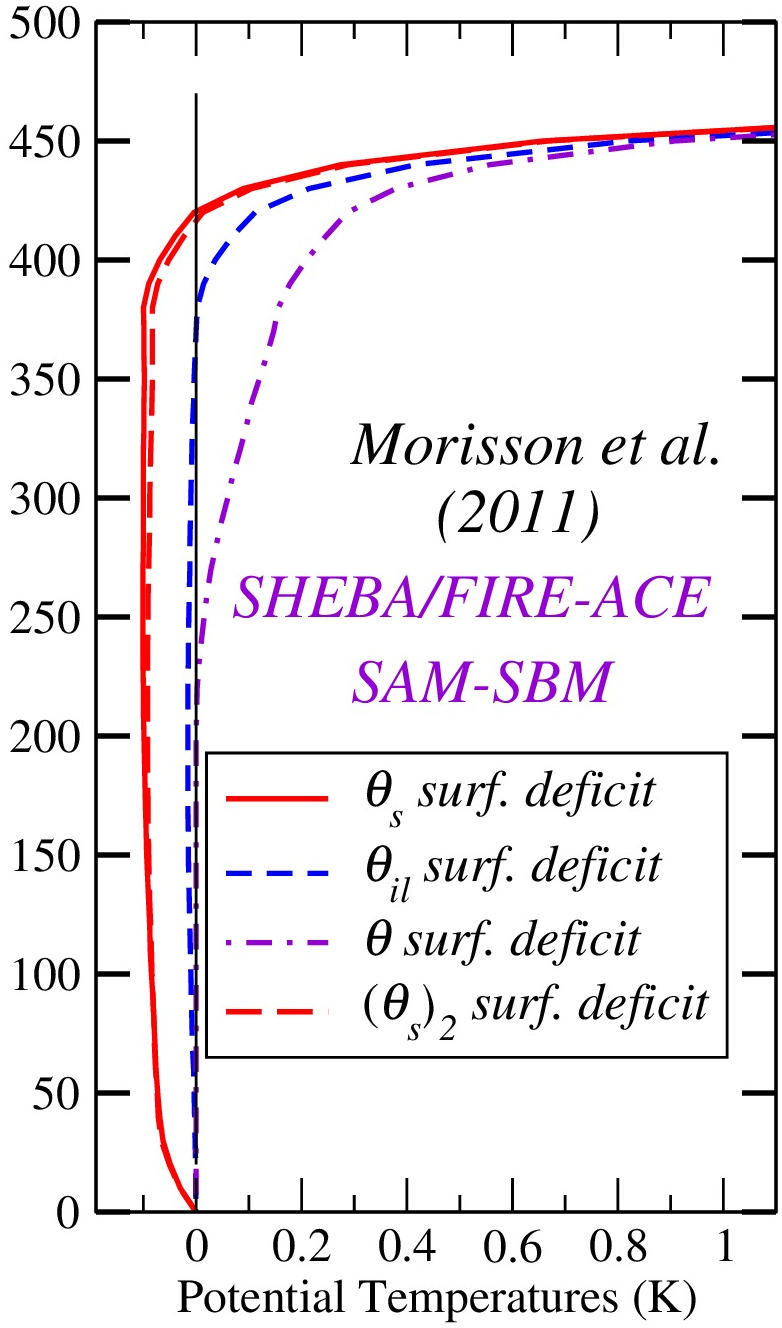}
\vspace{-3mm}
\caption{\small \it 
Same as in Fig.~\ref{fig_1} but for vertical profiles (left) and surface deficit values (right) for several potential temperatures: 
$\theta_s$ given by (\ref{eq_3})~; 
$\theta_{il}$ defined in Tripoli and Cotton (1981)~;
$\theta = T \: (p_0/p)^{(R_d/c_{pd})}$~;
the second-order value
$({\theta}_{s})_2$ defined in M15 but multiplied by the third line of (\ref{eq_3}).
\label{fig_3}}
\end{figure}
The paradigm for describing and simulating mixed-phase cloud is to consider that the ice-liquid potential temperature $\theta_{il}$ is a conservative variables, where $\theta_{il}$ defined in Tripoli and Cotton (1981) is similar to the bracketed terms in the first line of (\ref{eq_3}), except that the latent heats $L_v(T_0)$ and $L_s(T_0)$ are computed at the triple-point temperature $T_0 = 273.16$~K (not at $T$).


The conserved (namely constant) feature observed for $\theta_{il}$ in the PBL of Fig.\ref{fig_3} is likely due to the choice of the ice-liquid water static energy $h_L$ as a prognostic variables in the SAM-SBM runs, where $h_L = c_{pd} \: T + g \: z - L_v(T_0) \: q_l - L_s(T_0) \: q_s$ is clearly a proxy for $\theta_{il}$.

Differently, it is shown in Fig.\ref{fig_3} that the mixed-phase moist-air entropy value $\theta_s$ given by   (\ref{eq_3}) is not conserved (with $({\theta}_{s})_2$ being  indeed a good approximations of $\theta_s$).
This may be interpreted as an impact of the term $\exp ( \Lambda_r \: q_t )$ in the first line of (\ref{eq_3}) and due to changes in $q_t$ shown in Fig.\ref{fig_1} close to the ground (below $50$~m).


This impact of $q_t$ was missing in the definition of $\theta_{il}$ and in the approximate integration of the first and the second laws of thermodynamics derived in Dutton (1976, see before Eq.30, p.284, in the 1986 edition).

The ``equivalent'' version $\theta_{eil}$ defined in Tripoli and Cotton (1981) includes a factor $\exp [ \: ( L_v(T_0) \: q_t)/(c_{pd} \: T) \: ]$ which depends on $q_t$, where  $L_v(T_0)/(c_{pd} \: T) \approx 9$.
This factor is however different from the one  $\: \exp ( \Lambda_r \: q_t ) \:$ appearing in $\theta_s$ given by (\ref{eq_3}), where $\Lambda_r \approx 6 $ depends on the Third-Law reference values $(s_v)_r$ and $(s_d)_r$.
Only $\theta_s$ with $\Lambda_r \approx 6 $ is an {\it equivalent\/} of the moist-air entropy.


 \section{\underline{\Large Conclusions}} 
\label{section4}
\vspace{-4mm}

The search for ``conserved'' variables based on approximations of the moist-air entropy (function or equation) should be replaced by the use of the {\it conservative \/} variables  $\theta_s$ given by (\ref{eq_3}) which is a true {\it equivalent\/} variable.

A model using the mixed-phase version (\ref{eq_3}) for $\theta_s$ as a prognostic variable, including for turbulent and mass-flux mixing processes, could lead to more accurate results.
The impacts of the last two terms of (\ref{eq_3}) are to be investigated (ex. for supercooled water).

\vspace{1mm}
\noindent{\large\bf \underline {References}}
\vspace{0mm}

{\small 
\noindent{$\bullet$ Dutton J. A.} {(1986)}.
The Ceaseless Wind.
{\it Dover Publications}.
617 pages.
}

{\small \noindent{$\bullet$ Marquet P.} {(2011, M11)}.
Definition of a moist entropic potential temperature. 
Application to FIRE-I data flights.
{\it Q. J. R. Meteorol. Soc.}
{\bf 137} (656):
p.768--791.
}

{\small \noindent{$\bullet$ Marquet P.} {(2015, M15)}.
An improved approximation for the moist-air entropy potential temperature $\theta_s$.
{\it WGNE Blue-Book\/}.
\url{http://arxiv.org/abs/1503.02287}
}

{\small \noindent{$\bullet$ Morrison H. {\it et al.\/}} {(2011)}.
Intercomparison of cloud model simulations of Arctic 
mixed-phase boundary layer clouds observed during 
SHEBA/FIRE-ACE
{\it J. Adv. Model. Earth Syst. (JAMES)}
{\bf 3} (2): 
p.1942-2466.
}

{\small \noindent{$\bullet$ Shupe  M. D. {\it et al.\/}} {(2008)}.
Vertical Motions in Arctic Mixed-Phase Stratiform Clouds.
{\it J. Atmos. Sci.\/}
{\bf 65} (4): 
p.1304-1322.
}

{\small \noindent{$\bullet$ Tripoli G. J., Cotton  W. R.} {(1981)}.
The Use of lce-Liquid Water Potential Temperature 
as a Thermodynamic Variable In Deep Atmospheric Models.
{\it Mon. Wea. Rev.\/}
{\bf 109} (5): 
p.1094--1102.
}

  \end{document}